\documentclass[%
 reprint,
superscriptaddress,
 amsmath,amssymb,
 aps,
]{revtex4-2}

\usepackage{graphicx}
\usepackage{dcolumn}
\usepackage{bm}
\usepackage{xcolor}

\begin{document}

\preprint{APS/123-QED}

\title{Leveraging machine learning features for linear optical interferometer control}

\author{Sergei Kuzmin}
\email{s.kuzmin@rqc.ru}
\affiliation{Quantum Technologies Centre, Lomonosov Moscow State University, Russia, Moscow, 119991, Leninskie Gory 1 building 35}
\affiliation{Russian Quantum Center, Russia, Moscow, 121205, Bol'shoy bul'var 30 building 1}
\author{Ivan Dyakonov}%
\email{iv.dyakonov@quantum.msu.ru}
\affiliation{Quantum Technologies Centre, Lomonosov Moscow State University, Russia, Moscow, 119991, Leninskie Gory 1 building 35}
\affiliation{Russian Quantum Center, Russia, Moscow, 121205, Bol'shoy bul'var 30 building 1}
\author{Stanislav Straupe}%
\affiliation {Sber Quantum Technology Center, Kutuzovski prospect 32, Moscow, 121170, Russia}
\affiliation{Quantum Technologies Centre, Lomonosov Moscow State University, Russia, Moscow, 119991, Leninskie Gory 1 building 35}
\affiliation{Russian Quantum Center, Russia, Moscow, 121205, Bol'shoy bul'var 30 building 1}

\date{\today}

\begin{abstract}
We have developed an algorithm that constructs a model of a reconfigurable optical interferometer, independent of specific architectural constraints. The programming of unitary transformations on the interferometer's optical modes relies on either an analytical method for deriving the unitary matrix from a set of phase shifts or an optimization routine when such decomposition is not available. Our algorithm employs a supervised learning approach, aligning the interferometer model with a training set derived from the device being studied. A straightforward optimization procedure leverages this trained model to determine the phase shifts of the interferometer with a specific architecture, obtaining the required unitary transformation. This approach enables the effective tuning of interferometers without requiring a precise analytical solution, paving the way for the exploration of new interferometric circuit architectures.
\end{abstract}

\maketitle


\section{\label{sec:Introduction}Introduction}

Linear optical interferometers have become essential instruments in quantum optics \cite{Carolan2015} and optical information processing \cite{Harris2018}. The growing interest in the field is driven by the increased availability of integrated photonics technology. A distinguishing characteristic of modern interferometers is their reconfigurability, which enables them to modify their impact on input optical modes as needed. This feature particularly enhances their appeal for information processing tasks. Reconfigurable interferometers play a pivotal role in linear optical quantum computing experiments \cite{Wang2018, Wang2019, Zhang2021} and are considered as potential accelerators for deep learning applications \cite{Hamerly2019, Wetzstein2020}. Moreover, high fabrication quality and improved scalability have led to the development of field-programmable photonic arrays, which are adaptable circuits suitable for a broad range of applications via low-level programming \cite{PrezLpez2020}.

The transformation matrix $U$ defines the operation of a linear interferometer, relating inputs to outputs as $a^{(out)}_{j}=\sum_{i}U_{ij}a^{(in)}_{i}$. This transformation is controlled by parameters $\{\varphi\}$, such as phase shifters, expressed as $U = U(\{\varphi\})$. An interferometer is termed universal if it can implement any unitary $N\times N$ matrix. Programming the interferometer involves adjusting the parameters $\{\varphi\}$ to achieve the desired matrix $U$. A well-established approach is the Hurwitz decomposition \cite{Hurwitz1897}, realized with two-port Mach-Zehnder interferometers (MZI) \cite{Reck1994, Clements2016}. Although this mesh architecture is straightforward to program, it demands perfect balance in the MZIs, and optimization methods are used to address imperfections, which complicates the programming process \cite{Burgwal2017, Dyakonov2018}.

Addressing the impact of fabrication defects has led to the development of more complex architectures \cite{Saygin2020, Fldzhyan2020, Fldzhyan:24, Kondratyev:24}, which resist straightforward analytical description and require the use of optimization methods for programming. Implementing the transformation in a physical device demands an experimental, resource-heavy matrix reconstruction process \cite{Tillmann_2016} for each iteration of the selected optimization algorithm. This process is impractical for users, as it requires re-optimization whenever transformation needs change. First, optimizing within high-dimensional parameter spaces is time-intensive, demanding intricate adjustments without guaranteeing a global minimum. Second, rapid convergence in multi-parameter optimization algorithms typically relies on gradients, with gradient estimation precision being constrained by measurement noise. Lastly, while the number of phase-shifter switching cycles is technically unlimited, prolonged optimization degrades the programmable circuit's lifespan.

Recent research has made significant progress in tackling this issue. For example, \cite{Kuzmin:21} proposes a machine learning approach to train a numerical model of a linear optical interferometer. This method resembles supervised learning, but simplifies the problem to non-convex optimization. Additionally, algorithms based on linear algebra have been developed to reconstruct these structures without the need for optimization procedures \cite{Bantysh:23, Bantysh_2024}.

In our study, we present two methods for programming linear optical interferometers with complex internal structures. We use feature engineering techniques on linear optical circuits and demonstrate that, in these defined features, the interferometer model is linearized. Consequently, its training can be generally performed using linear algebraic methods, avoiding the non-convex optimization required in \cite{Kuzmin:21}. This approach is particularly effective for shallow interferometers. The interferometer model is trained on a variety of transformations representing different phase shifts. This trained model facilitates rapid determination of the necessary phase shifts for a given unitary transformation through optimization applied to the model rather than the physical device itself. Our learning algorithm consists of two stages, similar to \cite{Kuzmin:21}: training the interferometer model based on a set of transformation examples, and programming, which involves determining the model's phase shifts for the desired transformation. Additionally, we explore the use of an {\it alternating least squares} (ALS) \cite{Pearson01111901} optimization technique on the physical interferometer. This technique is suitable for deep circuits, where it combines tomography-based and model learning-based approaches.

\section{\label{sec:Mathematical foundations}Mathematical foundations}

\begin{figure}[!t]
\includegraphics[width=1.0\linewidth]{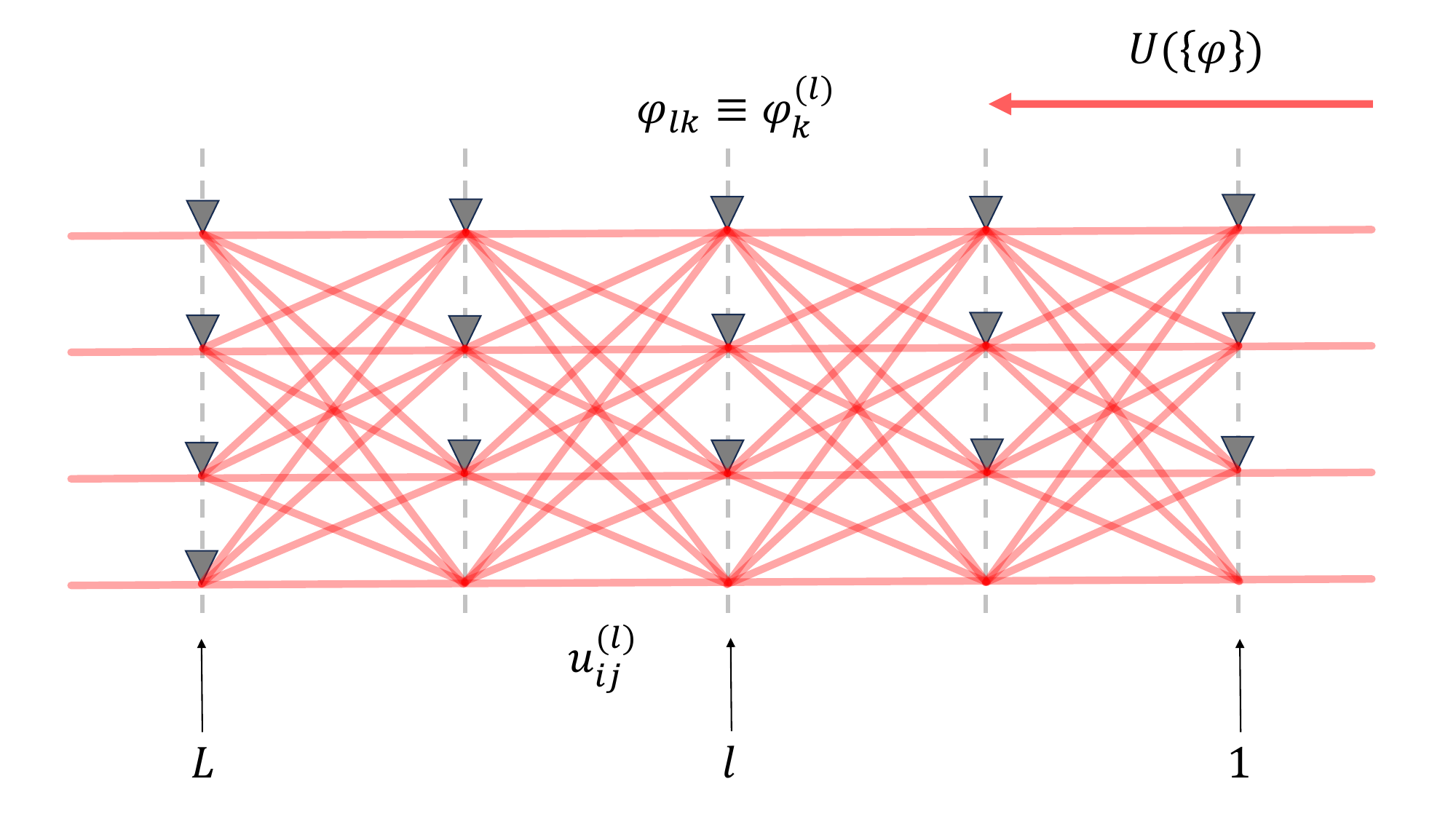}
\caption{\label{fig:linear_network} The schematic of the multimode interferometer structure and its integrated photonic implementation circuit. The basis matrices $U_{\ell}$ describe how light is distributed between the interferometer channels. The corresponding integrated elements can be implemented as waveguide lattices, where the waveguides are coupled, allowing energy transfer between different waveguides \cite{Skryabin:21}.}
\end{figure}

Our algorithm is developed to tackle the challenge of programming a multimode interferometer, which is composed of alternating phase-shifting and mode-mixing layers. This architecture is demonstrated to be nearly universally applicable, yet it lacks a simple correlation between the matrix elements and the interferometer's phase shifts \cite{Saygin2020}. It serves as an ideal example to showcase the core functionality of our algorithm. The circuit topology is illustrated in Fig. \ref{fig:linear_network}.

The unitary matrix $U$ is represented as:
\begin{equation}\label{Expansion}
    U = \Phi_L U_{L - 1} \dots U_{\ell} \Phi_{\ell} \dots \Phi_2 U_1 \Phi_1,
\end{equation}
where $\Phi_{\ell} = \operatorname{diag}\{ e^{i\varphi_{\ell k}}\} = \operatorname{diag}\{ e^{i\varphi_{k}^{(\ell)}} \}$ and $\{ U_{\ell} \}_{ij} = u_{ij}^{(\ell)}$. For further purposes, we introduce the notation $e^{i\varphi_{k}^{(\ell)}} \overset{\text{def}}{\equiv} \phi_{\ell k} \overset{\text{def}}{\equiv} \phi_k^{(\ell)}$. Thus, $\{ \Phi_{\ell} \}_{ij} = \delta_{ij} e^{i\varphi_{\ell j}} = \delta_{ij}e^{i\varphi_j^{(\ell)}} = \delta_{ij} \phi_j^{(\ell)}$.

We term $U_{\ell}$ as the {\it basis} matrices because they definitively characterize the interferometer's function. Once $U_{\ell}$ is known, a straightforward numerical optimization determines the required phase shifts $\varphi_{\ell k}$, completing the device's programming. Notably, the generalized expansion form (\ref{Expansion}), as discussed in \cite{Saygin2020}, applies universally to any linear-optical interferometer design. It can be readily demonstrated that any optical interferometer, comprising independent components such as fixed beamsplitting elements of any topology and phase modulators, can be decomposed into a sequence of unitary transformations that couple the circuit modes and the phase shifters. The critical information necessary includes the number of mode mixing and phase shifting layers. The flexibility inherent in the beamsplitting component structures and the absence of constraints on the length, sequence, or quantity of mode mixers and phase shifters underpin our algorithm's architecture-agnostic nature.

\section{\label{sec:The learning algorithm}The learning algorithm}

\subsection{\label{sec:all_layers_learning}All-layers learning}

Consider equation (\ref{Expansion}) expressed element-by-element, particularly the interferometer matrix element $u_{ij}$:

\begin{equation}\label{Convolution}
    u_{ij} = \sum_{k_1, \dots, k_{L - 2}} \phi_i^{(L)} u_{i k_1}^{(L - 1)} \phi_{k_1}^{(L - 1)} \dots u_{k_{L - 2} j}^{(1)} \phi_j^{(1)}.
\end{equation}

Next, we segregate the factors related to the basis matrices $u_{pr}^{(\ell)}$ from those related to phase shifts $e^{i\varphi_{k}^{(\ell)}} \overset{\text{def}}{\equiv} \phi_k^{(\ell)}$:

\begin{equation}\label{Convolution_Transposition}
    u_{ij} = \sum_{k_1, \dots, k_{L - 2}} \underbrace{u_{i k_1}^{(L - 1)} \dots u_{k_{L - 2} j}^{(1)}}_{w_{i \bar{k} j}} \underbrace{\phi_i^{(L)} \phi_{k_1}^{(L - 1)} \dots \phi_j^{(1)}}_{\theta_{i \bar{k} j}}.
\end{equation}

\begin{figure}[!t]
\includegraphics[width=1.0\linewidth]{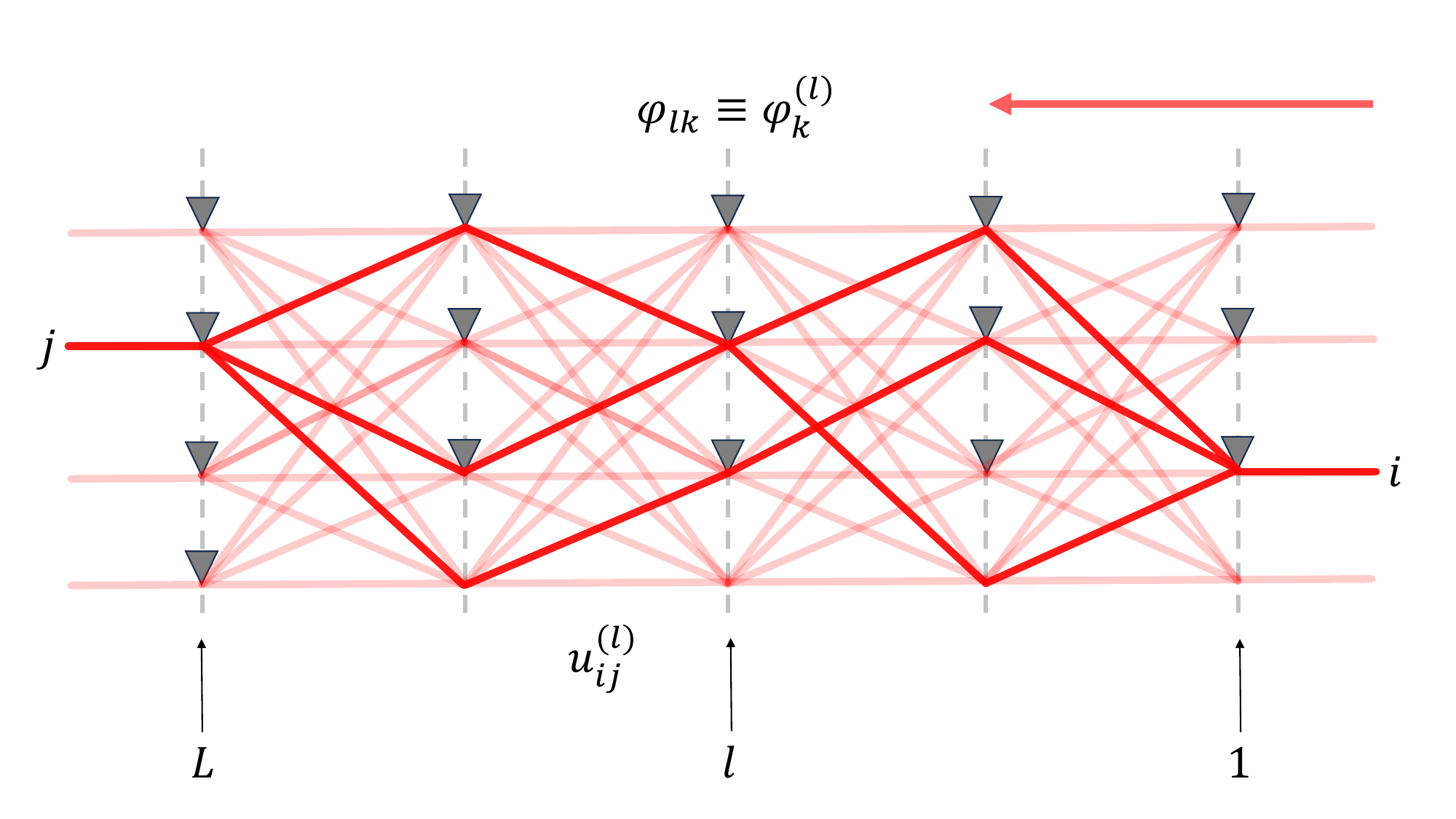}
\caption{\label{fig:feed_forward_calc} Illustration of the calculation of the matrix element $u_{ij}$ according to equation (\ref{Convolution_Transposition}). The matrix element $u_{ij}$ is formed by summing the products of weights $w_{i \bar{k} j}$ and new features $\theta_{i \bar{k} j}$ over all paths connecting input mode $i$ and output mode $j$. The figure shows for example the summation over only three paths for fixed indices $i$ and $j$, but in general the summation is over all paths. The weights $w_{i \bar{k} j}$ are composed of products of the basis matrix elements $u_{p r}^{(\ell)}$, while the features $\theta_{i \bar{k} j}$ represent the product $e^{i\varphi_{k}^{(\ell)}} \overset{\text{def}}{\equiv} \phi_k^{(\ell)}$.}
\end{figure}

We observe that our model is linearizable, with each element of the interferometer matrix $u_{ij}$ represented as a scalar product between a weight vector $w_{i \bar{k} j}$ and a feature vector $\theta_{i \bar{k} j}$, which depend solely on the phase shifts:

\begin{equation}\label{Linear_Regression}
    u_{ij} = \sum_{\bar{k}} w_{i \bar{k} j} \theta_{i \bar{k} j}, \hspace{0.5cm} \bar{k} = (k_1, \dots, k_{L - 2}).
\end{equation}

Thus, the dependence of the linear-optical interferometer matrix U on the phase shifts $\varphi_{k}^{(\ell)}$ can be characterized as a linear model using these features, with the third-rank tensor $w_{i \bar{k} j}$ comprehensively describing this model.

Let $i, j$ be fixed, then $u_{ij} \overset{\text{def}}{\equiv} u$, $w_{i \bar{k} j} \overset{\text{def}}{\equiv} w_{\bar{k}}$, and $\theta_{i \bar{k} j} \overset{\text{def}}{\equiv} \theta_{\bar{k}}$:
\begin{equation}\label{Simple_Linear_Regression}
    u = \sum_{\bar{k}} w_{\bar{k}} \theta_{\bar{k}}.
\end{equation}

Suppose we have $M$ measurements $u^{(m)}$ at $M$ different phase setups $\varphi_{k}^{(l)}$, corresponding to $M$ values $\theta_{\bar{k}}^{(m)}$. This results in a system of linear equations:
\begin{equation}\label{Matrix_Linear_Regression}
    \Theta W = U,
\end{equation}
where the new features $\theta_{\bar{k}}$ are taken into account. The matrix $\Theta$ serves as an object-feature matrix, where each of its 
$M$ rows represents a measurement, and each column corresponds to features $\theta_{\bar{k}}$. The vector $W$ consists of the weights $w_{\bar{k}}$ arranged in a column vector format, while $U$ is a column vector composed of $u^{(m)}$ values.

The system of linear equations (\ref{Matrix_Linear_Regression}) for the experimental tomography data of the linear-optical interferometer $U$ also includes random noise $\varepsilon$:

\begin{equation}\label{Matrix_Linear_Regression_Noisy}
    \Theta W = U + \varepsilon.
\end{equation}

To correctly solve such a problem, the columns of the object-feature matrix must be linearly independent. Additionally, the number $M$ of rows -- representing the different sets of phase shifts $\{\varphi_{k}^{(\ell)}\}$ used for the tomography of the transformation matrices $U^{(m)}$ of the linear-optical interferometer -- should be at least equal to the number of features. This setup allows the problem of finding the weight vector (for fixed indices $i$ and $j$) to be formulated as a least-squares problem:

\begin{equation}\label{Loss_Func}
    ||\Theta W - U||_{2}^2 \rightarrow \underset{\text{W}}{\operatorname{min}}.
\end{equation}

Such a minimization problem can be solved analytically using Moore-Penrose pseudoinversion:

\begin{equation}\label{Moore_Penrose}
    W = (\Theta^{\dagger}\Theta)^{-1}\Theta^{\dagger} U.
\end{equation}

We now consider a general functional (\ref{Loss_Func}) that incorporates the indices $i$ and $j$:

\begin{equation}\label{Loss_Func_Tensor}
    \mathcal{L} = \dfrac{1}{M} \sum_{m = 1}^{M} \sum_{i, j}^{N} \Bigl(\sum_{\bar{k}} w_{i \bar{k} j} \theta_{i \bar{k} j}^{(m)} - u_{ij}^{(m)}\Bigr)^2 \rightarrow \underset{w_{i \bar{k} j}}{\operatorname{min}}.
\end{equation}

The matrix $F = (\Theta^{\dagger}\Theta)^{-1}\Theta^{\dagger}$, as shown in equation (\ref{Moore_Penrose}), remains constant for each index $i$ and $j$ in the matrix element $u_{ij}$. Therefore, it is necessary to calculate $F$ only once, after which we can directly articulate a solution for the tensor $w_{i \bar{k} j}$ that minimizes the functional $\mathcal{L}$ given by equation (\ref{Loss_Func_Tensor}):

\begin{equation}\label{weights_analytically}
    w_{i \bar{k} j} = \underset{w_{i \bar{k} j}}{\operatorname{argmin}} \hspace{0.1cm} \mathcal{L} = \sum_{m} F_{\bar{k}m} u_{ij}^{(m)}.
\end{equation}

The problem of minimizing the functional (\ref{Loss_Func_Tensor}) can also be addressed numerically, for example, by employing {\it gradient descent} (GD) or {\it stochastic gradient descent} (SGD) methods.

Let's now discuss the data requirements for training. For the algorithm to function effectively, the number of rows $M$ in the matrix $\Theta$ from equation (\ref{Moore_Penrose}) -- that is, the number of phase sets $\{\varphi_{k}^{(\ell)}\}$ where we perform tomography on the interferometer transformation matrix $U^{(m)}$ -- must be at least equal to the number of columns. This number is dictated by the new features $\theta_{i \bar{k} j} = \phi_i^{(L)} \phi_{k_1}^{(L - 1)} \dots \phi_j^{(1)} = e^{i\varphi_{i}^{(L)}} e^{i\varphi_{k_1}^{(L - 1)}} \dots e^{i\varphi_{j}^{(1)}}$. In the case of a fully connected linear optical network, as depicted in Figures \ref{fig:linear_network} and \ref{fig:feed_forward_calc}, the number of these features is $\mathcal{O}(N^{L})$, resulting in inefficient scaling with increasing mode number $N$, assuming the circuit $L \sim N$ is complete. Therefore, this algorithm is most effective when the network has a small depth $L \ll N$.

So far, we have addressed the scenario where a physical linear-optical interferometer is used to derive its digital model (\ref{Linear_Regression}) through a weight tensor $w_{i \bar{k} j}$ by performing tomography on its transformation matrices $U^{(m)}$ at different phase shifts. This is followed by optimizing the phase delays $\varphi_{k}^{(\ell)}$ in the digital model to achieve the desired target transformation $U_{\operatorname{target}}$. However, we can also consider a different approach -- optimizing our physical interferometer layer by layer.

\subsection{\label{sec:layer_wise_learning}Layer-wise learning}

\begin{figure}[!t]
\includegraphics[width=1.0\linewidth]{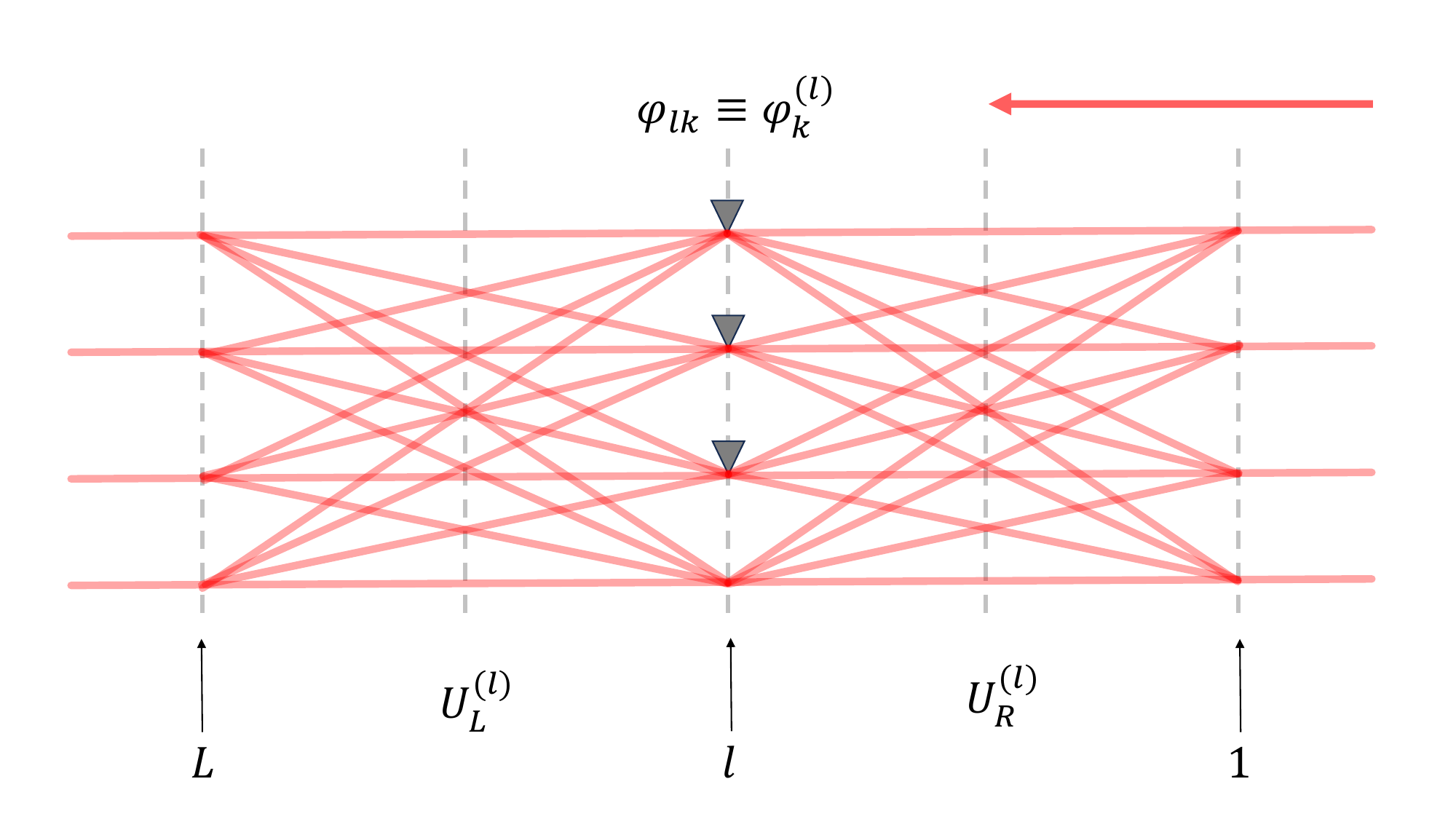}
\caption{\label{fig:one_layer} Illustrating the approach of considering a single phase layer of a linear-optical interferometer, we consider static basis matrices $U_{L}^{(\ell)}$ and $U_{R}^{(\ell)}$ (\ref{One_Layer}) to the left and right of the phase layer at each step of the algorithm, although they depend on the corresponding phase shifts.}
\end{figure}

To obtain the target transformation $U_{\operatorname{target}}$, we proceed through the interferometer layer by layer, possibly iterating over the layers multiple times sequentially. At a given layer $\ell$, we adjust only the phases in this layer while keeping others constant. This approach represents an implementation of the {\it alternating least squares} (ALS) method for our scenario. By considering only the current phase layer $\ell$ as the variables to be optimized, we can locally represent our interferometer during training as follows:
\begin{equation}\label{One_Layer}
    U = U_{L}^{(\ell)} \Phi^{(\ell)} U_{R}^{(\ell)},
\end{equation}
i.e., as a single layer. This is demonstrated in Fig. \ref{fig:one_layer}. Importantly, this approach is also equivalent to the model described in (\ref{Linear_Regression}), although in this scenario, our features are simplified to merely the phase exponents of $\theta_{i \bar{k} j} = \phi^{(\ell)}_k = e^{i \varphi_{k}^{(\ell)}}$. Subsequently, we can train a linear model for this single layer using $\mathcal{O}(N)$ tomographic samples and numerically optimize the phases to closely approximate the target transformation $U_{\operatorname{target}}$:
\begin{equation}
    \Phi^{(\ell)} = \underset{\Phi^{(\ell)}}{\operatorname{argmin}}||U_{\operatorname{model}}(\Phi) - U_{\operatorname{target}}||_{2}^2.
\end{equation}

This numerical optimization problem can be solved using methods such as gradient descent or the BFGS algorithm \cite{Liu1989}. After performing this local minimization, we proceed to the next layer, conduct tomography, and train a local linear model (\ref{Linear_Regression}).

The advantage of this approach is that the sample size for training scales linearly as $\mathcal{O}(N)$. However, a disadvantage is the need to perform tomography on a physical device at each layer. In contrast, the previous algorithm required many tomographies only once to reconstruct the full model of the linear-optical interferometer. Here, we don't need to directly measure the phase gradient of the loss function on the device, as it is derived from the local linear model of the interferometer.

\section{\label{sec:Numerical experiment}Numerical experiment}

In this section, we present the main performance measures of the interferometer model learning and tuning algorithms presented in Section \ref{sec:all_layers_learning} and Section \ref{sec:layer_wise_learning}.

\subsection{\label{sec:sim_all_layers_learning}Simulation of all-layers learning}

Here, we investigate the characteristics of the algorithm introduced in Section \ref{sec:all_layers_learning}, concentrating on the linear-optical interferometer shown in Fig. \ref{fig:linear_network}. To model the interferometer, we generate $L - 1$ basis matrices 
$U_{\ell}$, which are randomly sampled from the uniform Haar distribution via QR decomposition \cite{mezzadri2006generate}.

The task of the all-layers learning algorithm is to reconstruct the complete model of the linear-optical interferometer (\ref{Linear_Regression}) from a training dataset $(\{\varphi^{(m)}\}, U^{(m)})$ of size $M$. To create this dataset, we uniformly generate random phases between $0$ and $2\pi$, defining each $U^{(m)}$ using the generated basis matrices, these random phases, and the decomposition (\ref{Expansion}).

In real physical systems, when obtaining the dataset $(\{\varphi^{(m)}\}, U^{(m)})$, there are also tomography errors $\varepsilon$ (\ref{Matrix_Linear_Regression_Noisy}) present, which we must consider in our modeling. To account for such errors, we add random noise to each element $u^{(m)}_{ij}$ of the matrices $U^{(m)}$:

\begin{equation}
    \{U^{(m)}_{\operatorname{noise}}\}_{ij} = u^{(m)}_{ij} + \varepsilon^{(m)}_{ij},
\end{equation}
which we obtain in this way:

\begin{equation}
    \varepsilon^{(m)}_{ij} = \dfrac{\varepsilon}{\sqrt{2}}(x + iy),
\end{equation}
where $x \in \mathcal{N}(0, 1)$, $y \in \mathcal{N}(0, 1)$ are random variables and $\varepsilon$ is a fixed parameter. Thus, we obtain an updated dataset $(\{\varphi^{(m)}\}, U^{(m)}_{\operatorname{noise}})$ that also contains Gaussian noise. Then, when the training dataset is generated, we train the model weights $w_{i\bar{k}j}$ analytically using the formula (\ref{weights_analytically}).

To test the quality of training, we select random phases $\{\varphi\}$  that are not part of the training set $(\{\varphi^{(m)}\}, U^{(m)}_{\operatorname{noise}})$. We then evaluate the model's prediction for these phases, $U_{\operatorname{model}}(\{\varphi\})$, and compare it to the true values $U(\{\varphi\})$ (ignoring noise).  We use the Frobenius norm as the distance metric between $U_{\operatorname{model}}(\{\varphi\})$ and $U(\{\varphi\})$:

\begin{equation}
    \mathcal{L}_{\operatorname{Frobenius}} = \dfrac{1}{N}||U_{\operatorname{model}}(\{\varphi\}) - U(\{\varphi\})||^2_2.
\end{equation}

To investigate the properties of the learning algorithm, we use averaged results based on the Frobenius metric across many samples of linear-optical interferometers, training datasets, and test evaluations.

\begin{figure}[!t]
\includegraphics[width=1.0\linewidth]{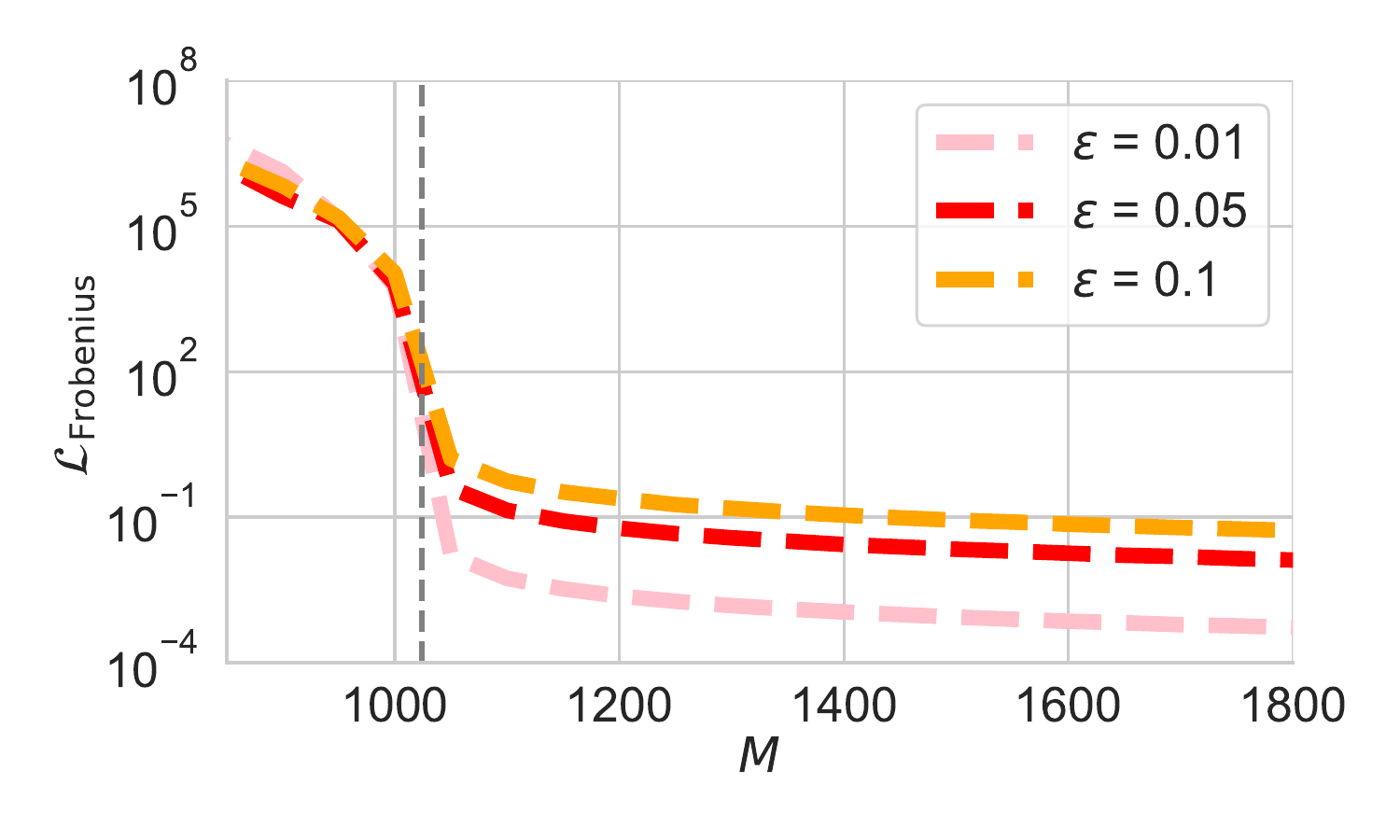}
\caption{\label{fig:learning_frobenius_w_noise_threshold} Dependence of the Frobenius metric on the test set as a function of the training sample size $M$. Averaging is performed over $1000$ examples of linear-optical interferometers $N = 4$ modes and $L = 5$ layers. The gray vertical dashed line indicates the minimum required training sample size for successful validation on the test set, $M_{\min} = 1024$.}
\end{figure}

Fig. \ref{fig:learning_frobenius_w_noise_threshold} shows the dependence of the Frobenius metric, averaged over $1000$ samples of linear-optical interferometers, on the size $M$ of the training sample for different noise levels $\varepsilon$, with $N = 4$ modes and $L = 5$ layers. We observe that successful learning (indicating high performance on the test set) occurs only beyond a certain minimum threshold $M_{\min}$, as discussed in Section \ref{sec:all_layers_learning}.

\begin{figure}[!t]
\includegraphics[width=1.0\linewidth]{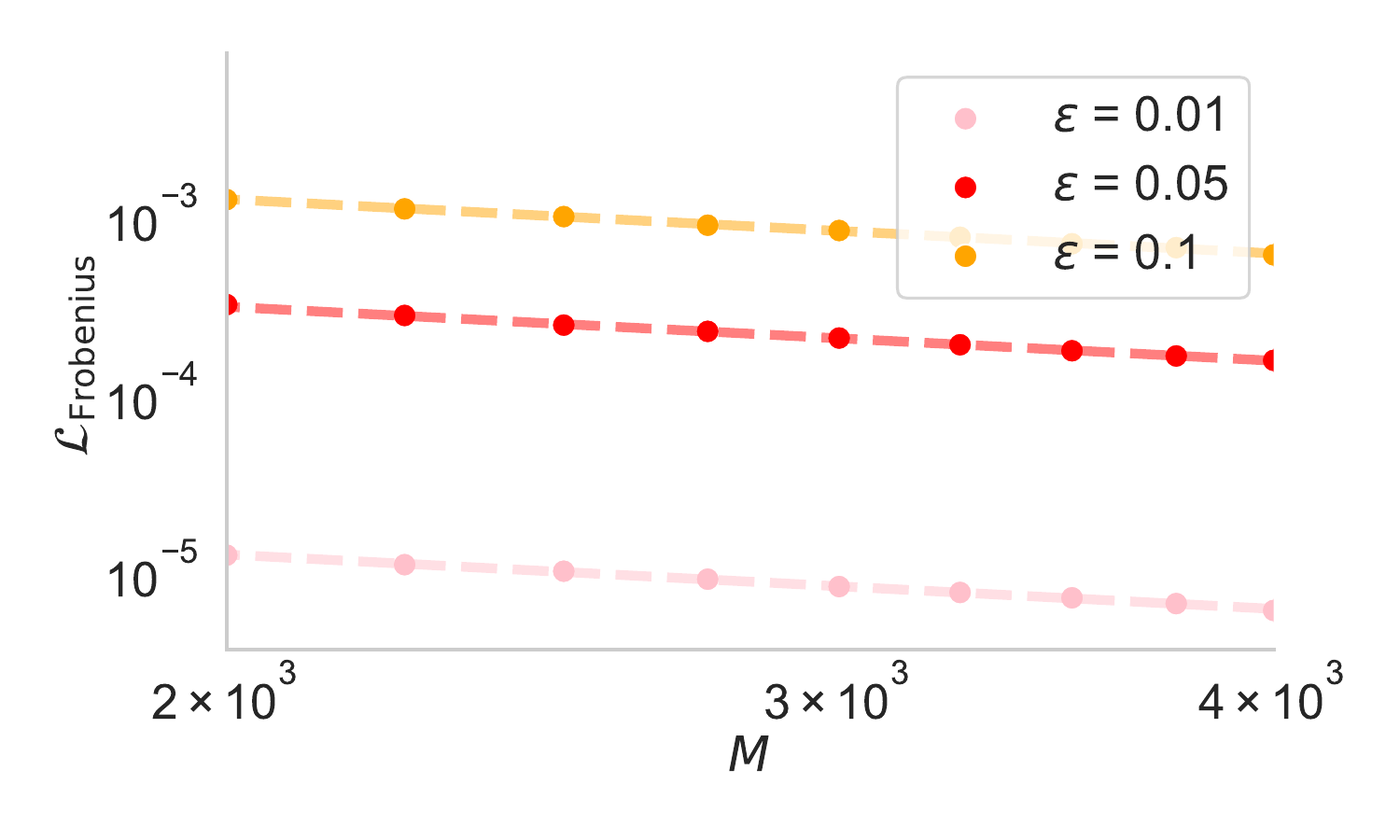}
\caption{\label{fig:learning_frobenius_w_noise_after_threshold} The dependence of the Frobenius metric on the test case as a function of the training sample size $M$ . Averaging is performed on $1000$ examples of linear-optical interferometers with number of modes $N = 3$ and number of layers $L = 4$. Values considered are $M \gg M_{\min} = 81$. Simulation points are approximated by the relation $\mathcal{L}_{\operatorname{Frobenius}} = \mathcal{C}N^{-k}$ using the least squares method, with results: $k = (1.017 \pm 0.022)$ for $\varepsilon = 0.01$, $k = (1.010 \pm 0.013)$ for $\varepsilon = 0.05$, $k = (1.014 \pm 0.024)$ for $\varepsilon = 0.1$.}
\end{figure}

\begin{figure}[!t]
\includegraphics[width=1.0\linewidth]{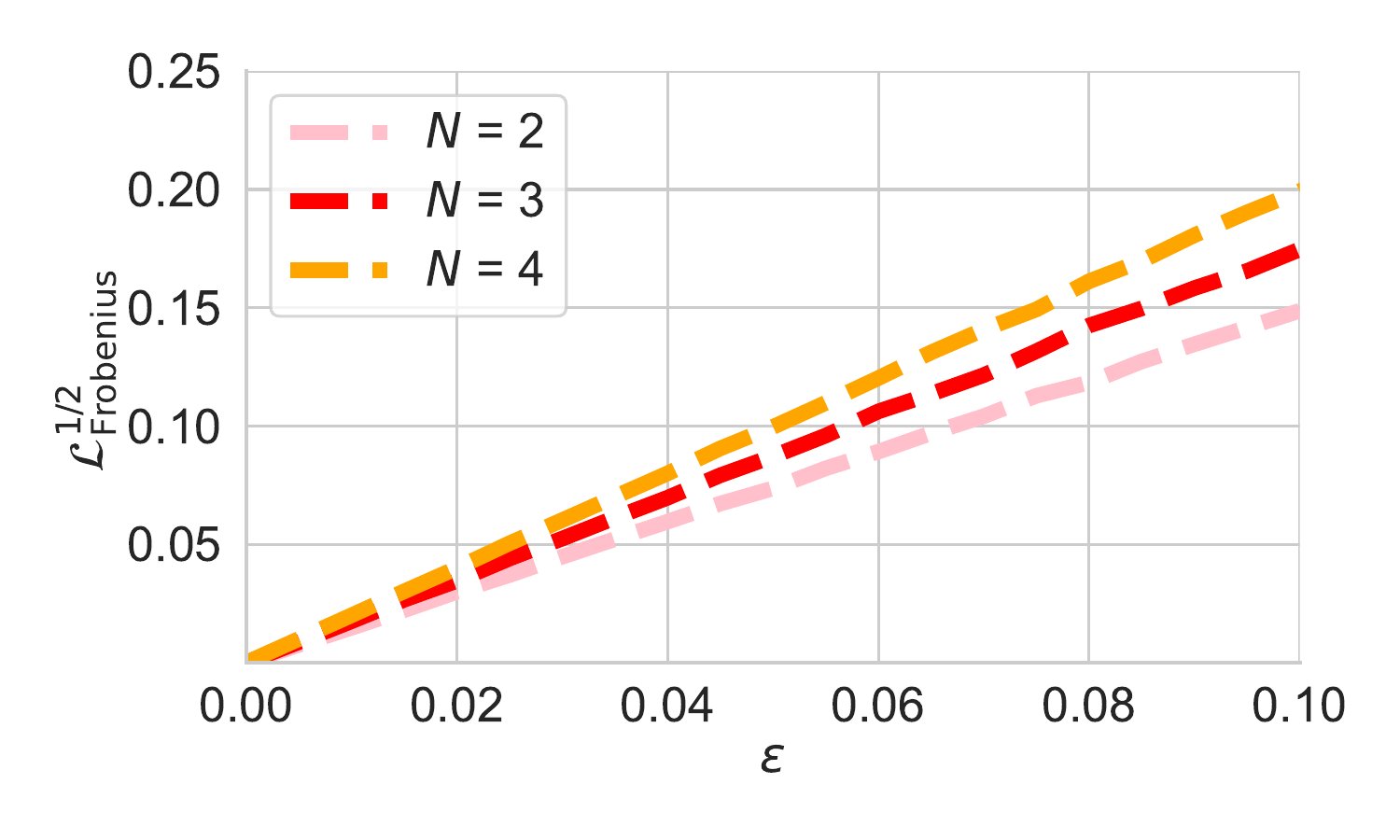}
\caption{\label{fig:learning_frobenius_w_noise_eps} Dependence of the Frobenius metric on the test case as a function of the noise level $\varepsilon$ in the training sample. Averaging is performed over $10000$ examples of linear-optical interferometers for $N = 2$ and $1000$ examples for $N = 3$ and $N = 4$. Full-layer interferometers, where $L = N + 1$, are considered. The training sample size is fixed at twice the minimum sample size, $M = 2 M_{\min}$, with $M$ equals $16$, $162$, and $2048$, respectively.}
\end{figure}

It is interesting to study how this dependence behaves asymptotically at large $M$. It is logical to assume that the Frobenius metric will decrease inversely proportional to $M$ as the variance of the sample mean, i.e. $\mathcal{L}_{\operatorname{Frobenius}} \sim \mathcal{O}(M^{-1})$. This is confirmed by Fig. \ref{fig:learning_frobenius_w_noise_after_threshold}, where the dependence of $\mathcal{L_{\operatorname{Frobenius}}}$ on $M$, at values significantly exceeding the threshold $M_{\min}$, is linear on a log-log scale. The least-squares approximation of the numerical plot by the dependence 
$\mathcal{C}M^{-k}$ provides an estimate for $k$ around 1. In Fig. \ref{fig:learning_frobenius_w_noise_eps}, the dependence of the square root of the Frobenius metric $\mathcal{L}_{\operatorname{Frobenius}}$ on the noise magnitude $\varepsilon$ is shown. This trend appears linear.

\subsection{\label{sec:sim_layer_wise_learning}Simulation of layer-wise learning}

In this subsection, we focus on investigating the properties of the layer-wise learning algorithm described in Section \ref{sec:layer_wise_learning}. We evaluate the properties of the interferometer tuning process concerning a specific transformation $U_{\operatorname{target}}$, considering the number of interferometer modes $N$ and its depth $L$.

We also focus on the architecture of a linear-optical interferometer, depicted in Fig. \ref{fig:linear_network}. Initially, we generate the $L - 1$ basis matrices $U_{\ell}$, which are sampled randomly from the Haar-uniform distribution via QR decomposition \cite{mezzadri2006generate}.

Subsequently, we generate random phases uniformly distributed between $0$ and $2\pi$, and use these to define $U_{\operatorname{target}}$. This involves utilizing the generated basis matrices, the random phases, and the decomposition (\ref{Expansion}) to establish $U_{\operatorname{target}}$. At this point in the simulation, we fix our linear-optical interferometer and the target transformation $U_{\operatorname{target}}$ that we aim to implement.

Further, the task of the interferometer tuning algorithm is to begin with an initial approximation of phases $\{\varphi\}$. After several iterations of the algorithm described in Section \ref{sec:layer_wise_learning}, the goal is to find phases such that the interferometer transformation $U$ closely approximates the target transformation $U_{\operatorname{target}}$, as measured by the Frobenius norm: $||U_{\operatorname{model}}(\{\varphi\}) - U_{\operatorname{target}}||_{2}^2 \rightarrow \underset{\{ \varphi \}}{\operatorname{min}}$.

We perform $1000$ iterations through the layers of the linear-optical interferometer to execute the ALS algorithm. In each iteration, the algorithm sequentially processes the layers of the interferometer $\ell$ from $1$ to $L$. Within each layer, tomography is simulated for $M = N + 1$ phases (as we must have at least $N$ training samples) while phases in other layers remain fixed. For each sample, the model (\ref{One_Layer}) is trained as detailed in Section \ref{sec:all_layers_learning}. Using this trained model, the algorithm minimizes the functional using the BFGS algorithm (ALS step), with $5$ iterations. In practical applications, the elements of the training sample set $(\{\varphi^{(m)}\}, U^{(m)})$ are derived from $U^{(m)}_{rec}$ unitary reconstruction algorithms \cite{Tillmann_2016, Suess2020}, applied to a reconfigurable interferometer programmed with $\{\varphi^{(m)}\}$ phases. The complexities of experimentally assembling a suitable training set are discussed in \ref{sec:Discussion}.

\begin{figure}[!t]
\includegraphics[width=1.0\linewidth]{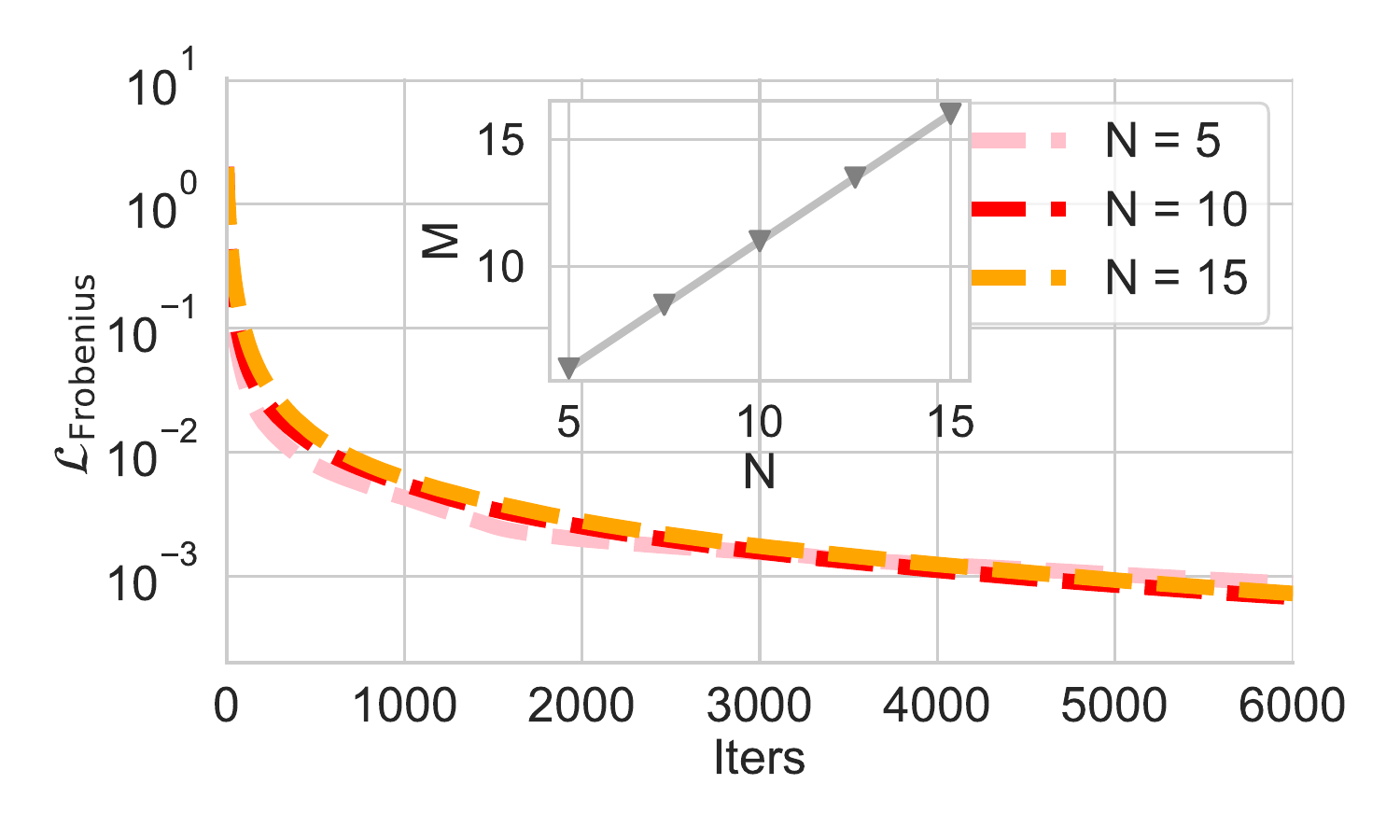}
\caption{\label{fig:tuning_frobenius} Illustration of the tuning process for a full-layer ($L = N + 1$) linear-optical interferometer. The figure demonstrates the dependency of the Frobenius loss function, averaged over 50 distinct runs of the tuning algorithm $\mathcal{L}_{\operatorname{Frobenius}} = \dfrac{1}{N}||U_{\operatorname{model}}(\{\varphi\}) - U_{\operatorname{target}}||_2^2$, on the number of iterations, defined as the product of layer passes and the total layers in the interferometer. The inset graph depicts the linear relationship between the training sample size $M$ and the number of modes $N$ ($M = N + 1$).}
\end{figure}

We conducted an experiment with 50 different randomly generated linear-optical interferometers to observe how the Frobenius loss function $\mathcal{L}_{\operatorname{Frobenius}} = \dfrac{1}{N}||U_{\operatorname{model}}(\{\varphi\}) - U_{\operatorname{target}}||_2^{2}$ varies with the iteration number for a full-depth ($L = N + 1$) interferometer. The averaged tuning results of the linear-optical interferometer are illustrated in Fig. \ref{fig:tuning_frobenius}. Note that the required number of training examples $M$ at each fixed layer in the experimental tuning increases as $\mathcal{O}(N)$.

\begin{figure}[!t]
\includegraphics[width=1.0\linewidth]{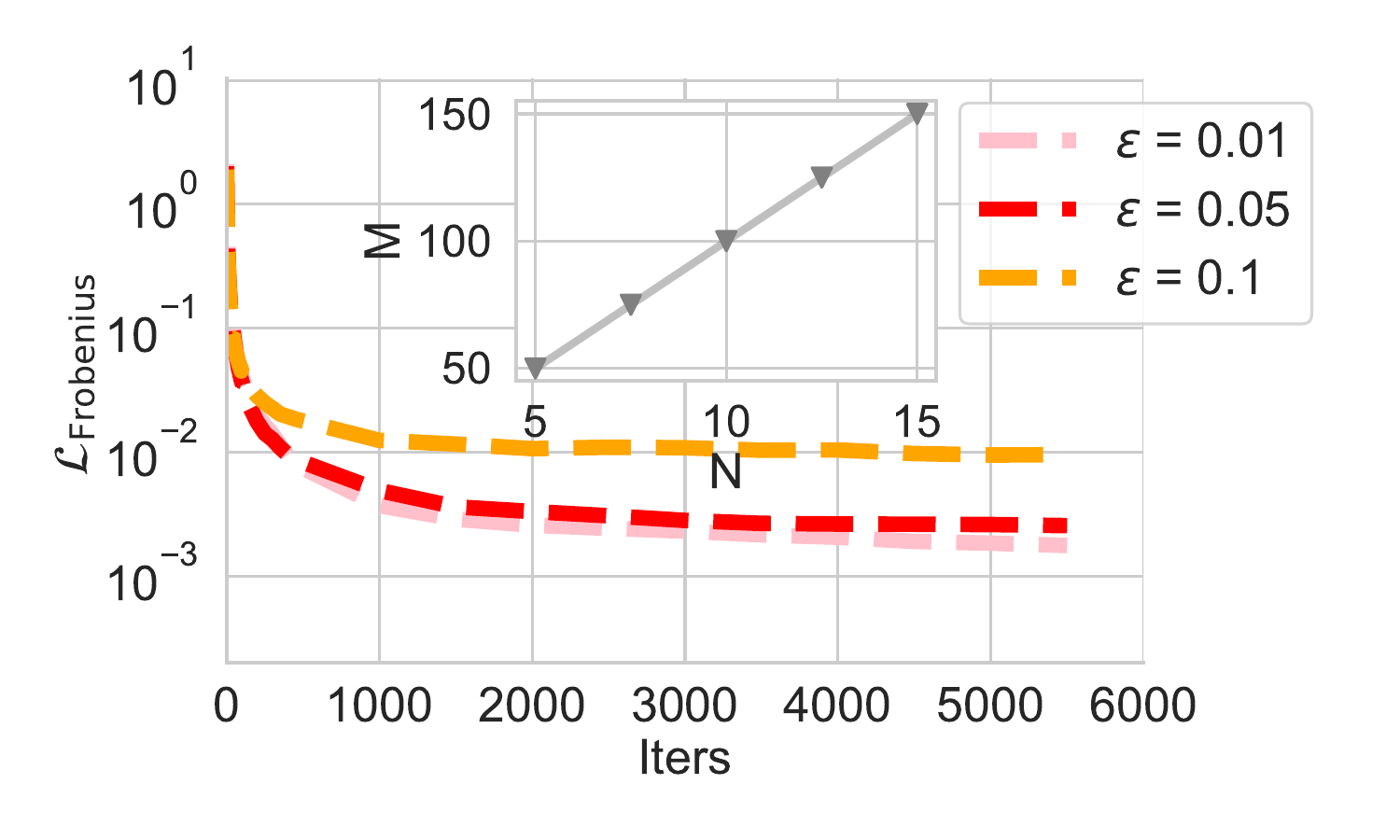}
\caption{\label{fig:tuning_frobenius_w_noise} Illustration of the tuning process for a full-layer ($L = N + 1$) linear-optical interferometer with layer-by-layer tomography executed under $\varepsilon$ noise conditions. This analysis considers $N = 5$, $L = 6$. The figure shows how the Frobenius loss function, averaged across 50 runs of the tuning algorithm $\mathcal{L}_{{\operatorname{Frobenius}}} = \dfrac{1}{N}|||U_{\operatorname{model}}(\{\varphi\}) - U_{\operatorname{target}}|||_2^2$, varies with the number of iterations, where each iteration is defined as the product of layer passes and total layers. The inset plot illustrates the linear relationship of the training sample size $M$ with the number of modes $N$ ($M = 10 N$).}
\end{figure}

Fig. \ref{fig:tuning_frobenius} shows the results of tuning the linear-optical interferometer without taking into account the noise $\varepsilon$ and random errors occurring during the layer-by-layer tomography process. In real devices, of course, there is noise, modeling of which we discussed in Section \ref{sec:sim_all_layers_learning}. We will also take noise into account in the process of interferometer tuning. For this purpose, we will do the same as for Fig. \ref{fig:tuning_frobenius}, except that now in each layer the tomography is modeled for $M = 10N$ phases (since according to the results in Fig. \ref{fig:learning_frobenius_w_noise_threshold} and Fig. \ref{fig:learning_frobenius_w_noise_after_threshold}) the model training error decreases with increasing $M$) and taking into account the $\varepsilon$ noise, as described in Section \ref{sec:sim_all_layers_learning}. The results of such tuning with noise are shown in Fig. \ref{fig:tuning_frobenius_w_noise}. Note that the required number of training samples $M$ on each fixed layer still increases as $\mathcal{O}(N)$ during the experimental tuning.

\section{\label{sec:Discussion}Discussion}

The demonstrated approach to interferometer programming stands out with several major advantages. First and foremost the method is agnostic of the architecture of the interferometer. Any universal interferometers reported in literature \cite{Reck1994, Clements2016, Saygin2020, Fldzhyan2020, Fldzhyan:24} admit some form of the expansion (\ref{Expansion}) -- the optical mode mixing elements interleaved with phase shifters. This means that both the gist of the algorithm and the mathematical framework fit any architecture of choice. These assumptions remain valid only if the mode mixers and the phase shifters are considered as independent elements.

Note that we propose two new methods for tuning linear-optical interferometers. The first method, detailed in Section \ref{sec:all_layers_learning}, applies to shallow schemes where $L \ll N$ \cite{Fldzhyan:24}. Its advantage lies in performing a one-time tomography procedure on the given physical linear-optical interferometer to determine transformations $U^{(m)}$ for different phase settings $\{\varphi^{(m)}\}$. This allows for the training of a comprehensive model of the interferometer (\ref{Linear_Regression}), enabling the determination of $U_{\operatorname{target}}$ for specific phase configurations. The second method, outlined in Section \ref{sec:layer_wise_learning}, is suitable for schemes of any depth, including those where $L \sim N$. However, it requires optimization on the physical device each time a new transformation $U_{\operatorname{target}}$ is desired. Despite this, it facilitates the layer-by-layer training and numerical optimization of the model (\ref{One_Layer}) to find optimal phases.

The bottlenecks of the proposed algorithm are related to the experimental issues. The $\mathcal{L}$ Frobenius metric (\ref{Loss_Func_Tensor}) requires exact measurement of the unitary elements' modulus and phase. Several reconstruction methods have been proposed and verified \cite{Laing2012, RahimiKeshari2013, Tillmann_2016, Spagnolo2017, Suess2020}. Some of them \cite{Laing2012, Suess2020} provide only partial information about the transformation matrix of the interferometer omitting phases which are impossible to reconstruct using the method-specific dataset. Any method will inevitably suffer from the path-dependent optical loss which is impossible to distinguish and attribute to the particular path inside the circuit. Another issue which is not covered by our algorithm arises from the crosstalks between the phase shifters. Our framework assumes that the phases in different paths are enabled independently which is not the case due to the crosstalks between different phase modulating elements. Luckily the integrated photonic modulator implementations typically exhibit extremely low crosstalks \cite{Zhang2020, Jiang2018}.

We believe that our results will enable opportunities to employ new programmable optical interferometer architectures for both classical and quantum applications. 

\section{\label{sec:Acknowledgements}Acknowledgements}
ID acknowledges support from the Russian Science Foundation grant 22-12-00353-P (https://rscf.ru/en/project/22-12-00353/). We are grateful to M.Y.~Saygin and S.A.~Fldzhyan for enlightening discussions.

\bibliography{bibtex}

\nocite{*}


\end{document}